\documentclass[manuscript,screen]{acmart}
\AtBeginDocument{%
  \providecommand\BibTeX{{%
    \normalfont B\kern-0.5em{\scshape i\kern-0.25em b}\kern-0.8em\TeX}}}

\setcopyright{acmcopyright}
\copyrightyear{2023}
\acmYear{2023}
\acmDOI{XXXXXXX.XXXXXXX}

\usepackage{graphicx}
\usepackage{tikz}
\usetikzlibrary{shapes.multipart, decorations.pathmorphing}
\usetikzlibrary{decorations.pathreplacing}
\usetikzlibrary{shapes, arrows}
\usepackage{xcolor}

\usepackage{url}
\usepackage{booktabs,makecell, multirow, tabularx}
\usepackage[export]{adjustbox}
\usepackage[belowskip=0pt,aboveskip=0pt]{caption}

\usepackage{mathtools,xparse}
\usepackage{multirow}

\usepackage{float}
\usepackage{subcaption}
\usepackage{epstopdf}
\usepackage{amsthm} 
\usepackage{amsfonts}
\usepackage{thmtools}
\usepackage{thm-restate}
\DeclareUnicodeCharacter{2212}{-}
 
\declaretheoremstyle[%
  spaceabove=6 pt,%
  spacebelow=6 pt,%
  bodyfont=\itshape,
  headfont=\normalfont \scshape ,%
  postheadspace=0.2em,%
  qed=\qedsymbol,%
  headpunct = .
]{mystyle}

\declaretheoremstyle[%
  spaceabove=6 pt,%
  spacebelow=6 pt,%
  bodyfont=\normalfont,
  headfont=\normalfont \scshape ,%
  postheadspace=0.2em,%
  qed=\qedsymbol,%
  headpunct = .
]{proofstyle}

\declaretheorem[name={Proof},style=proofstyle,unnumbered,]{Proof}
\declaretheorem[name={Definition},style=mystyle,numbered,]{Definition}
\usepackage{proof-at-the-end}

\usepackage{hyperref}
\usepackage{cleveref}

\usepackage{fancyhdr}
\usepackage{amsmath}
\usepackage[linesnumbered,ruled,vlined,noend]{algorithm2e}

\def\0{\varnothing}

\def\Su2mN{\sum_{i=2}^{N}}
\def\Su2m{\sum_{i=2}^{N}}

\def\.{\mathcal{\cdot}}

\def\1{{\boldsymbol 1}}

\begin{document}

\title{Attack Detection Using Item Vector Shift in Matrix Factorisation Recommenders}

\author{Sulthana Shams}
\affiliation{%
  \institution{Trinity College Dublin}
  \country{Ireland}}
\email{sshams@tcd.ie}

\author{Douglas Leith}
\affiliation{%
  \institution{Trinity College Dublin}
  \country{Ireland}}
\email{Doug.Leith@tcd.ie}


\begin{abstract}
 This paper proposes a novel method for detecting shilling attacks in Matrix Factorization (MF)-based Recommender Systems (RS),
in which attackers use false user-item feedback to promote a specific item. Unlike existing methods that use either use supervised learning to distinguish between attack and genuine profiles or analyse target item rating distributions to detect false ratings, our method uses
an unsupervised technique to detect false ratings by examining shifts in item preference vectors that exploit rating deviations and
user characteristics, making it a promising new direction. The experimental results demonstrate the effectiveness of our approach in
various attack scenarios, including those involving obfuscation techniques.
\end{abstract}






\maketitle

\section{Introduction}

In Recommender Systems (RS), users are grouped based on their interests which motivates many advertising strategies \cite{clustering_ads_1,clustering_ads_2,clustering_ads_3}. Recent research proposes exploiting such clustering of users in RS for their ’hiding in the crowd' privacy benefits \cite{BLC,FloC,Fledge}. Unfortunately, despite privacy benefits, user clusters in RS are also susceptible to data poisoning attacks \cite{segment_attack_analysis, TORS}. For example, social media platforms have received attention for discriminatory and predatory targeting of user communities \cite{facebook_ads, ads_google}. Further, adversaries have been known to use RS to influence users by introducing false product reviews \cite{amazon_fake_reviews} to promote misinformation. 

The vulnerability of RS to potential manipulation by malicious users was initially investigated by \cite{profile_injection_attacks,fun_and_profit,segment_attack_analysis}. These attacks involve injecting false profiles to either promote or demote specific items and are called Data Poisoning/Shilling Attacks in RS. Push attacks elevate targeted items, while nuke attacks seek to demote them. Such fraudulent ratings and profiles can severely undermine the RS robustness. We assume that the attacker can only inject a limited number of fake users. Each fake user rates a limited number of items (including the target item and other non-target items called filler items) to evade suspicion. To detect these attacks, various approaches have been explored over the years. For example, exploiting features of attack profiles to distinguish “attack” profiles from “regular” profiles \cite{defending_attacks_1,defending_attacks_2}, assigning reputation scores to users in RS \cite{manipulation_resistant_RS} and analysing rating distributions of items in time  \cite{item_time_series_1, time_series_2} are some of the approaches explored to make RS robust.

In this work we re-visit the robustness problem to attacks in a Matrix Factorisation (MF) based RS \cite{MF}.
MF is widely known in RS due to its simplicity and effectiveness. The typical paradigm of MF in RS is to decompose the user-item interaction matrix $R \in \mathbb{R}^{n \times m}$ as the product of two low-dimensional latent matrices $U \in \mathbb{R}^{d \times n}$ and $V \in \mathbb{R}^{d \times m}$ such that their product $U^TV$ is approximating $R$. Matrix $U$ captures the relationship between a user and the latent features while matrix $V$ captures the relationship between an item and the features. We call $U$ as the user-feature matrix and $V$ as the item-feature matrix. 
Items are recommended in such an MF-based RS based on the proximity of the item vectors to the user vectors in the latent space. When a targeted attack, such as a push attack, is launched on a particular item, it will cause a shift of the item vector in the low-dimensional space, impacting the item's recommendation \cite{TORS}. Success of the attack depends on how much it can shift the item vector distribution significantly. Based on this observation, we propose a detection approach that examines the shift in item vectors.

We know that the effect of fake ratings is largest in groups, while individual fake ratings have negligible effects, especially in small numbers \cite{user_cluster_1}. As a result, eliminating clusters of fake ratings rather than individual ones makes sense.  If the feature vector of an item changes significantly after a block of ratings enters the RS, we can suspect that the ratings are fake and avoid using them to train the $U, V$ matrices. Many literary works assume that the target attack rating is always the highest value on a scale. An attacker can avoid detection based on such assumptions by simply providing a rating value one step lower \cite{obfuscated_attacks}. 
Our detection method significantly improves the identification of these obscured attack tactics highlighting the importance of considering rating variability in attack detection methods.  

In this paper, we present a novel detection method that is based on the deviation in the item vector and takes into account an item's overall preference information. With only $20-25$ true ratings per user cluster required during training, the experimental results achieve a superior detection precision over existing state-of-the-art detection strategies.

\section{Related Work}
Various shilling detection methods have been proposed that defend by detecting and removing fake profiles in RS, focusing mainly on extracting the signatures of authentic user profiles. The generated profiles will deviate statistically from those of authentic users and several attributes for detecting these anomalies are examined in \cite{feature_profile,defending_attacks_1, defending_attacks_2}. But a major obstacle in user focussed detection method is that it can recognize the distinctive signature of only the known attack model. For newer or hybrid attack models, the detection methods may fall short. 

In \cite{user_cluster_1}, authors proposed a PCA-based (Principal Component Analysis based) detection method that exploit the similarity structure in fake user profiles to separate them from normal user profiles using unsupervised dimensionality reduction methods. While the approach works well for attacks that show a good correlation among attack profiles, \cite{user_cluster_challenge} discusses effective attack strategies dropping the assumption of unusually high correlations among malicious attack profiles. They demonstrate that dimensionality reduction-based detection methods cannot detect such low-diversity attacks accurately.

Detecting shilling attacks in a model-free approach involves identifying abnormalities in the rating distribution of items as discussed in \cite{defending_attacks_3,time_series_2, item_time_series_1, rating_deviation, MPE}. 
Consider, \cite{defending_attacks_3,time_series_2, item_time_series_1} which analyses the changes to items in a time series. They investigate time intervals of rating activity that may suggest an item is under attack. They assume that attack profiles are injected in a short time and detection result is influenced by the selected time interval size. For example, \cite{time_series_2} uses item anomaly detection based on sample average and sample entropy in a time series. However, their method is tested only on dense items with at least $500$ ratings.  
Studies such as  \cite{rating_deviation,MPE}  look at the deviation of target item ratings from predicted ratings to identify abnormal ratings. 
Other works such as \cite{MF_robust, MF_robust_trim, manipulation_resistant_RS,adv_survey} aim to build manipulation-resistant RS to limit the damage from injected fake ratings. 

To the best of our knowledge, no previous work has investigated leveraging item preference vectors to detect attacks in MF-based RS. Our detection approach provides the next step by performing well against  obfuscated and unobfuscated attacks while requiring little amount of training data.

\section{Item Vector Shift Based Detection Model}
\subsection{User Cluster Based Recommendation Model}
In a MF-based RS, the user-item rating matrix $R$ is first factorized approximately as $U^TV$, where matrices $U$ and $V$ have relatively small inner dimensions $d$.  Each column in matrix $U$ is a length $d$ weight vector that captures a user's preferences, and each column in matrix $V$ is a length $d$ weight vector that captures the characteristics of one item. 

The standard approach to obtain $U, V$ given user-item ratings is to minimize the sum of squared error over the set of all observed ratings, i.e. 

\begin{equation}\label{costfunc_recap}
 \min_{U,V}\sum_{(i,j)\in \mathcal{O}} (R_{i,j}-U_i^TV_j)^2  + \lambda\left(\sum_{i}\lVert U_i\rVert^2+\sum_{j}\lVert V_j\rVert^2 \right)
\end{equation}

Where $\mathcal{O}$ is the set of (user, item) rating pairs, $R_{i,j}$ is the rating of item $j$ by user $i$, $U_i$ and $V_j$ are the column vectors that describe the preferences associated to user $i$ and item $j$ respectively; $\lVert.\rVert$ is the Euclidean norm and $\lambda$ is a regularisation weight. The predicted rating matrix given by the inner product $U^TV$ gives the individual predicted rating per user per item.  We deviate from the standard setup by assuming that users are part of separate groups. Each user belongs to one of the user groups and each group's members share similar preferences. To create the matrix $\tilde{U}$ of group vectors, we cluster the $U_i$ values. In our study, we use $k$-means to group columns of matrix $U$ such that users with similar preferences (i.e., weight vectors) are clustered together. Other clustering algorithms, such as the strategy in \cite{BLC} may also be used. Let $\tilde{U}_g \in \mathbb{R}^{d\times1}$ for $g=1,\cdots,|\mathcal{G}|$ be the preference vector associated with a group of users and gather these features to form matrix $\tilde{U} \in \mathbb{R}^{d \times |\mathcal{G}|}$. Each column in $\tilde{U}$ is a length $d$ weight vector that captures a group's preferences. For users belonging to a group $g$ with weight vector $\tilde{U_g}$ and items with weight vectors in $V$, the predicted rating matrix is the inner product $\tilde{U_g}^TV$, i.e., users belonging to the same group have the same predicted rating value per item.

\subsection{Utilizing Item Vectors for Improved Anomaly Detection}

Given a target item $j^*$ and its item vector $V_{j^*}$ based on initial rating information, we can recursively compute updates to $V_{j^*}$ \cite{TORS} \footnote{Proof of Theorem \ref{V_iterative} given in Appendix}. 
i.e. let $\hat{V}_{{j^*}}$ be the updated $V_{j^*}$ after receiving a block of $m$ new ratings, i.e. let a block of $m$ users with feature vectors in $X \in \mathbb{R}^{d \times m}$ and target item ratings in vector $y \in \mathcal{R}^{m \times 1}$ enter RS. Given $A^{-1}=\left(\sum_{i \in \mathcal{U}(j^*)}U_iU_i^T+\lambda I\right)^{-1}$ where  $\mathcal{U}(j^*)$ is the set of all true users who initially rated item $j^*$, we can write,
\begin{align}\label{V_iterative}
\hat{V_{j^*}}-V_{j^*}&=A^{-1}X\left(I+X^TA^{-1}X\right)^{-1}\left(y-X^TV_{j^*}\right)
\end{align}

The updated position of the item vector in the latent space given by equation \ref{V_iterative} depends on three factors: the initial user preference vectors of users who gave ratings (represented by matrix $A$), the feature values of users who provide the new ratings (represented by $X$), and the deviation factor (represented by $y - X^TV_{j^*}$).

The term $y - X^TV_{j^*}$ measures how much of the new rating vector ($y$) for the target item can be explained by the predicted rating ($X^TV_{j^*}$), which helps to distinguish between fake and real ratings. 
If $X^TV_{j^*}$ can already explain $y$, then the term is zero and $\hat{V_{j^*}}$ is equal to $V_{j^*}$. However, a significant deviation is anticipated when compared to ratings from actual users because fake ratings are typically set to the maximum or nearly maximum value on the rating scale. If there is no significant deviation, the target item already has a high rating and is not worth further attacking.  In addition to the deviation factor, the shift also depends on matrix $X$. The knowledge that the attacker has access to determines the feature vector values of attack profiles ($X$). Thus the shift to the item vector also depends on the information available to the attacker. 

To illustrate how attacks may shift the target item vector, suppose true users have $U=[1,-1]$ or $U=[-1,1]$.   i.e. the first set of users like items with $V=[1,0]$ and dislike items with $V=[0,1]$, but the second set of users are the opposite. Then an attack against the first set of true users keeping $U=[1,-1]$ constant to increase the rating of an item with $V=[0,1]$  may be performed by shifting $V=[0,1]$ to $V[1,0]$. Then such a shift, while it increases the rating in users with $U=[1,-1]$, would decrease the rating in group $U=[-1,1]$. So the attack effects leaks to all clusters due to the changes to item vector $V_{j^*}$. The approach that is adopted in this paper is to consider the consistency of the preference of user clusters to the target item vector.

Utilising item feature vectors for anomaly detection provides a distinct advantage over existing methods. Unlike other methods that rely solely on attack profile signatures or analysing item rating distributions, the item feature vectors capture the effect of both of these factors, promising a more robust and effective approach to detecting shilling attacks in RS.

%

\subsection{Proposed Item Vector Based Detection (IVD) Method}
\begin{figure}[H]
\centering
\subfloat[Effect of Increasing True Ratings on $V_{j^*}$]{
\includegraphics[width=0.50\columnwidth]{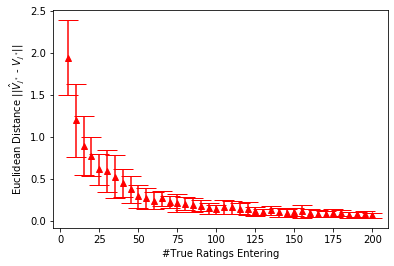}
}
\subfloat[$V_{j^*}$ Deviation from True,Fake Ratings]{
\includegraphics[width=0.50\columnwidth]{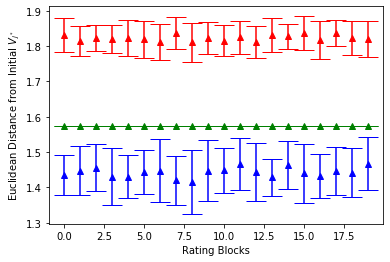}
}

\caption{Plot illustrating the IVD method in MovieLens 100k dataset}
\label{V}
\end{figure}

Consider that the target item $j^*$ initially received no ratings in the RS. Then, at random, we select a group and a user from that group to offer a rating.  The iterative change in distance (mean and standard deviation) between the updated $\hat{V}_{j^*}$ and the last updated $V_{j^*}$ after each block of true ratings is depicted in Figure \ref{V}(a). We see that the item vector shifts less from its previous value in response to new ratings after roughly $100$ in total true ratings, indicating that the item's preference has been established and that additional true ratings contribute little new preference information.  
 
Given a target item $j^*$, we randomly select a cluster $g$ as a reference in the d-dimensional vector space. We hypothesize that further real ratings will not significantly shift the vector $V_{j^*}$ in reference to cluster $g$, because true ratings rarely generate large changes in current preferences, as previously demonstrated. We suspect that the ratings are fraudulent if any block of new ratings leads the item vector to deviate significantly from reference cluster $g$. Significant deviations from the reference cluster may indicate that the new ratings are attempting to manipulate the recommendations by shifting the item vector into a different region of the vector space.

To demonstrate this concept, Figure \ref{V}(b) computes the distance ($D_{g,\hat{j}^*}$) between $\hat{V}_{{j^*}}$ to cluster $g$  and compares it to the distance ($D_{g,j^*}$) between initial $V_{j^*}$ and cluster $g$  for each false (indicated by red plot) 
and true rating blocks (indicated by blue plot). As we can see, each block of fake rating \footnote{modelled as Average Attack} results in a larger shift away from the initial distance $D_{g,j^*}$ (reference preference information: green plot), indicating a shift in preferences, whereas each block of real rating results in shifts relatively close to  the reference preference $D_{g,j^*}$.

This observation allows us to recognise and eliminate fake ratings from the RS. Note that the model can continue to recommend to fake users. Thus, only suspected ratings are removed before the RS retrains the $U, V$ to generate an updated prediction matrix. 

 \vspace{2mm}
\begin{algorithm}[H] 
\SetAlgoLined 
\KwIn{$V_{j^*}$, Cluster $g$, Reference Euclidean Distance $D_{g,j^*}(\tilde{U}_g, V_{j^*})$, Threshold =$th$}
 \For{ block index $n=1,2,3.\ldots$}{
  Find updated item vector {\bf  $\hat{V}_{j^*}$ } from eqn \ref{Uconst_iter}\\
  Calculate $D_{g,\hat{j}^*}$ between $\hat{V}_{j^*}$ and $\tilde{U}_g$;\\
  \eIf{$D{g,\hat{j}^*} > D_{g,j}+th$}
  {Remove the new block of ratings from the RS training process
       }  
   {Keep the new block of ratings for RS training process
  }}     
 \caption{Item Vector Based Detection Algorithm}\label{ivd}
\end{algorithm}

\section{Experiments and Evaluation Setup}

\subsection{Datasets}
We evaluate the effectiveness of attack on the MovieLens dataset (943 users rating 1682 movies, contains 100000 ratings from 1-5) which is widely used across literature for evaluating recommender systems under attack. Additionally we use larger Movielens dataset (6040 users, 3706 movies, contains 1 milion ratings from 1-5 ) and Netflix dataset. We take a dense subset of the Netflix dataset by selecting the top 1000 items and the top 10000 users who rated these items.

\subsection{Attack Model}
 
The attackers give high ratings to the target item to make it more visible, and a set of items are rated to create a fake user profile that mimics actual user behaviour. The rating distribution of these filler items determines the type of attack. We report results for two standard attack profiles mentioned in the literature: Random Attack and Average Attack \cite{fun_and_profit} in addition to another sophisticated attack type called Target Cluster Attack \cite{MF_ours} which is an extension of Average attack but aimed at specific user clusters. 

The \textit{Random Attack} is a zero-knowledge attack where ratings to filler items in each rating profile are distributed around the overall mean of items in the RS. The \textit{Average Attack} distributes the ratings for filler items in attack profiles around the mean for each item.  In the \textit{Target-Cluster Attack}, attackers target a specific user cluster \footnote{We randomly choose $g=2$ to target. Results are similar for any chosen $g$} by sampling filler ratings from a Gaussian distribution using the cluster-wise mean rating of items in the RS. The Average and Target attacks assume access to item mean ratings. This is a reasonable assumption as aggregate user preference information is often publicly available.

It's worth noting that we don't require mean ratings for every single item in the RS to execute these attacks. This is because most users only rate a small fraction of the entire product space, given the limitations of how many items a person can interact with. Therefore, attackers can focus on a subset of the product space to make their attacks effective \cite{segment_attack_analysis}. Ratings for popular items within the RS are relatively easy to obtain and can be used for these attacks. The popularity of these items can be determined using external data sources and doesn't necessarily depend on RS-specific data.

In the case of the target attack model, attackers can leverage general domain knowledge about item features to target users with preferences for specific types of items. For instance, they may spot products with genres that are comparable to the one they want to advertise and find possible targets by locating users who highly score these related items. In the RS, it is possible to then create fictitious profiles that closely resemble real ones by computing the mean ratings that this set of users provided for other items. 
Platforms like Amazon and Goodreads display user's ratings and reviews for items/books they've engaged with.  By filtering based on adversary's domain knowledge of the target item, an attacker can target user clusters likely to be pre-disposed to the targeted item. 

Moreover, the availability of public rating datasets has made finding preference ratings easier. For instance, a complete viewing history of any user may be acquired with limited adversary knowledge (2-8 ratings per user), which does not need to be precise, from the standard Netflix Dataset \cite{anonymity_break_1}. Similarly, the presence of overlapping sources of preference information can contribute to the identification of target users. For instance, Amazon could potentially re-identify users on competitors' websites by comparing their purchase history with reviews posted on those sites, and subsequently decide on marketing strategies for these customers. This idea of leveraging overlapping sources was explored and demonstrated by \cite{anonymity_break_2} using ML datasets and online forums.

Further, Facebook, Instagram and Amazon already offer targeted advertising/information sharing services where you can select any user segment based on preferences to the type of product you want to promote, age, location and other demographics \cite{awsblog, facebookbusiness}. 

\subsubsection{Target Item Ratings}

We consider items with ground truth average empirical rating $<3.5$ in all user clusters and with at least $20-25$ true ratings per user cluster, and attack profiles assign a maximum rating value to promote it in the RS. To make the detection harder, we also report results when using the target-shifting obfuscation technique \cite{obfuscated_attacks}, which involves shifting the rating given to the target item from the maximum rating to a rating one step lower.

Additionally, we also evaluate the effect of the choice of filler items. Choosing filler items appropriately makes the generated profiles similar to the genuine profiles making it harder to spot fake users.

\subsection{Baseline Detection Approach}
In a series of experiments, the presented method's detection performance was compared with several baselines. Here is a brief description of the methods used:

\begin{itemize}
    \item PCA-based: This method involves transforming the user-item matrix into a lower-dimensional space using Principal Component Analysis (PCA). Each user profile is then represented by three principal components. By analyzing the proximity of profiles to the new space's origin, potential attackers can be identified as users exhibiting suspicious behavior. We fix the $r=10\%$ i.e. the top $10\%$ of the users given by PCA is suspected as fake users. \cite{user_cluster_1}.
\item MPE-based: This method utilizes the mean prediction error (MPE) for individual items. It calculates the deviation between predicted ratings and actual ratings to identify anomalous rating behavior. If the item has significant MPE values, potential anomalies in the user ratings to the item can be detected \cite{MPE}. We fix the threshold as $1.5$ 
\end{itemize}

\subsection{Evaluation Metric}
We evaluate detection performance using two metrics: detection rate and false alarm rate. The detection rate is calculated by dividing the number of times of detected attack ratings by the total number of times fake rating blocks were inserted. The false alarm rate, on the other hand, is determined by the number of true rating blocks predicted as anomalies divided by the total number of true rating blocks inserted. The reported results are an average over $50$ randomly chosen target item from the set of potential target items. We set the threshold to 0.07 for IVD 

\section{Results and Discussion} \label{discussion}

In this section, we compare the performance of IVD against two state of the art detection strategies that reports a high detection rate and very low false alarm rate, namely PCA method \cite{user_cluster_1} and MPE method \cite{MPE}. Both methods are stable against attack and filler sizes with $>90\%$ detection rates as we will see later.

But both approaches have their limitations. For example, PCA works by 
observing that the profiles of shillers are very similar. They interpret users as variables (i.e. the dimensions of the data are the users, and the observations are the item ratings), then we have data where a number of dimensions are very similar and PCA can find a set of variables (users) which are highly correlated. Thus an attack profile that is undetectable by the PCA detector must reduce the differences from genuine profiles. The key factor here is in filler selection. Choosing filler items appropriately can bring down the correlation among fake users making them similar to other true users in the RS. When filler items are selected randomly, they are likely to have smaller pairwise overlaps compared to genuine users. Genuine users are more likely to rate certain items more frequently than others. 
Based on this observation,  study by \cite{user_cluster_challenge} has shown that an attacker should choose filler items according to their overall popularity among the genuine user base. A simple and effective strategy to obfuscate attacks is to choose filler items with equal probability from the top x\% of most popular items, rather than from the entire catalogue of items, to make detection harder. 

\begin{figure}[h]
    \centering
    
    \begin{subfigure}[b]{0.45\textwidth}
        \includegraphics[width=\textwidth]{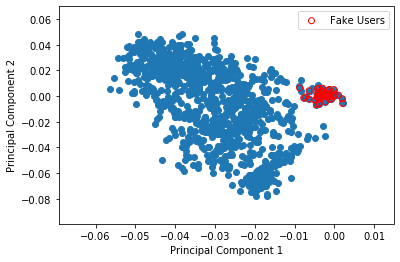}
        \caption{PC Space under Random Filler Choice}
        \label{fig:subfig1}
    \end{subfigure}
    \hfill
    \begin{subfigure}[b]{0.45\textwidth}
        \includegraphics[width=\textwidth]{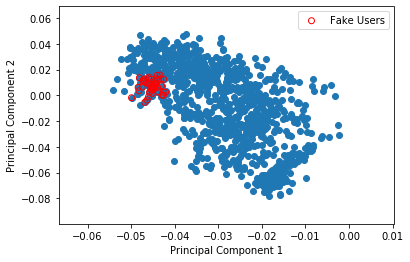}
        \caption{PC Space Under top $20\%$ Filler Choice}
        \label{fig:subfig2}
    \end{subfigure}
    
    \caption{Clusters in 2D Space for Normal Users and Fake Users (Avg Attack) in ML-100k}
    \label{fig:PCA}
\end{figure}

\begin{figure}[h]
    \centering    
    \begin{subfigure}[b]{0.45\textwidth}
        \includegraphics[width=\textwidth]{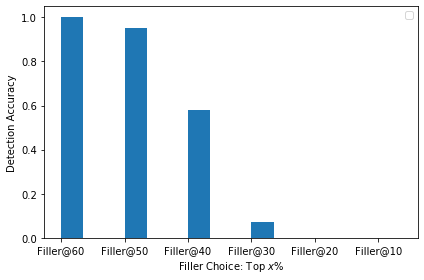}
        \caption{PCA Performance Against Filler Choice}
        
    \end{subfigure}
    \hfill
    \begin{subfigure}[b]{0.45\textwidth}
        \includegraphics[width=\textwidth]{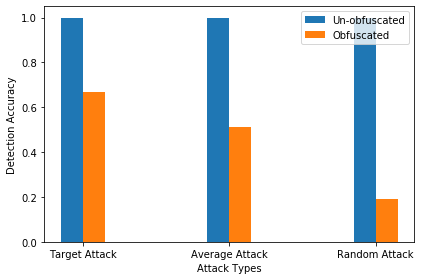}
        \caption{MPE performance against Target Rating Obfuscation}
        
    \end{subfigure}    
    \caption{Detection Accuracy for PCA and MPE in ML-100k}
    \label{fig:detection_pca_mpe}
\end{figure}

In Figure \ref{fig:PCA}, we plot the users in RS in the PC space. The coordinates here are the coefficients of each user in the 1st and 2nd principal component. Figure \ref{fig:PCA}(a) shows how, due to their low PC scores, the fake users are focused around the origin in the PC space. To identify fake users, PCA method assumes the two coefficents of each user represent a point in space and sort
the points in order of their distance from origin. The top $r\%$ eliminated users will contain fake users with more than $90\%$ accuracy.  The PC scores are difficult to distinguish when smaller subsets of popular item are used as filler items as can be observed from figure \ref{fig:PCA}(b) and thus fails to identify fake users. Figure \ref{fig:detection_pca_mpe}(a) shows how the detection rate of PCA against standard average attack. The detection rate decreases as the selection of most popular items is reduced to smaller subsets, with $0\%$ detection for $x\le20$. Although not demonstrated separately, we note that PCA shows no perceivable drop in detection accuracy in the face of target rating obfuscation as discussed in \cite{user_cluster_1}. This is because these are linear transformations, which are not very effective when linear dimensionality
reduction is performed.

MPE, on the other hand, is more concerned with deviations in item predicted ratings than with user profiles. It is relatively resistant to filler selection strategies and produced good results in our studies. The capacity of MPE to identify tiny changes in item predicted rating deviation made it a desirable tool for attack detection.  The assumption is that attackers just focus on the target item and rate it with the maximum or lowest rating many times in order to promote or demote the item to the recommendation list. The simple approach of obfuscating the target item rating from a maximum value to a value one step lower, on the other hand, leads the prediction error to be similar to that of the genuine user base and fails to detect the fake activity.

\begin{figure}[h]
    \centering
    
    \begin{subfigure}[b]{0.45\textwidth}
        \includegraphics[width=\textwidth]{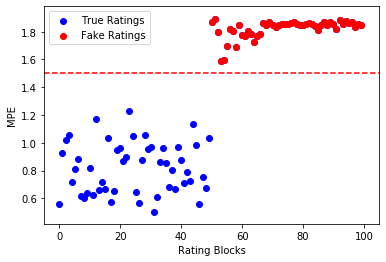}
        \caption{Target Item Rating: Maximum  }
        \label{fig:subfig3}
    \end{subfigure}
    \hfill
    \begin{subfigure}[b]{0.45\textwidth}
        \includegraphics[width=\textwidth]{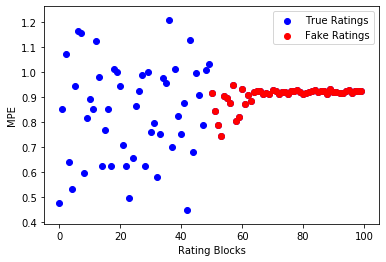}
        \caption{Target Item Rating: Obfuscated }
        \label{fig:subfig4}
    \end{subfigure}    
    \caption{Distribution of MPE for Genuine and Fake Rating Blocks in ML-100k}
    \label{fig:MPE}
\end{figure}

The MPE for average attack with and without target rating obfuscation approach is plotted in the figure \ref{fig:MPE}. Figure \ref{fig:MPE}(a) depicts how the fake ratings form a cluster away from true users due to their significant deviations. However, when target rating obfuscation is used, the MPE of false users becomes difficult to discern from that of legitimate users as is demonstrated in figure \ref{fig:MPE}(b). Figure \ref{fig:detection_pca_mpe}(b) compares the performance of MPE against standard attack model and when target rating is obfuscated for a fixed filler choice of top $20\%$. We can see that the detection rate decreases when compared to 100\% detection in un-obfuscated attacks. Thus, the choice of filler items has no effect on MPE, but the target rating obfuscation does since without obfuscation, MPE results in perfect detection for the $x=20\%$ filler choice. 

In the further sections, we compare how IVD performs under these obfuscation schemes to PCA and MPE approaches, and we give findings for three different datasets. We demonstrate that IVD performs more effectively than both techniques under these obfuscation strategies.

\subsection{Effectiveness of Attack Size and Filler Size}

\begin{figure}[H]
    \centering
    
    \begin{subfigure}[b]{0.45\textwidth}
        \includegraphics[width=\textwidth]{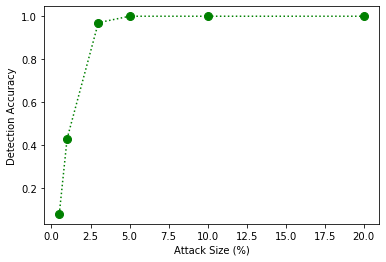}
        \caption{Detection Rate for varying attack sizes. Filler size=$10\%$}
    \end{subfigure}
    \hfill
    \begin{subfigure}[b]{0.45\textwidth}
        \includegraphics[width=\textwidth]{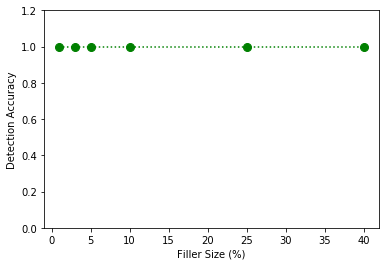}
        \caption{Detection Rate for varying filler sizes. Attack size=$5\%$}
    \end{subfigure}
    
    \caption{Effect of Attack and Filler Size IVD for ML-100k}
    \label{fig:attack_filler}
\end{figure}

In this experiment, we simulate an attack by introducing fake profiles alongside the regular user profiles in the recommender system. We examine how the performance of the IVD method  is affected by varying attack and filler sizes using the ML-100k dataset. The findings from this experimentation hold true for larger datasets such as Movielens 1M and Netflix. Our focus on the 100k ML dataset allows for direct comparison with previously reported results, as this dataset has been widely used to evaluate prior shilling detection methods.

We consider diverse attack sizes ($0.5\%$, $1\%$, $3\%$, $5\%$, $10\%$, $20\%$) and select the filler items randomly from the top $60\%$ of most popular items in the RS. We fix the number of filler items to $10\%$ of the total item count. In figure \ref{fig:attack_filler}(a), we observe that for attack size $<1\%$, IVD fails to detect the presence of fake ratings. Detecting an attack's impact is challenging when attack sizes are small. This is because such low attack sizes are unable to alter the item vector sufficiently to enhance the target item's rating. As a result, the attack's effect on the target item is insignificant. Because IVD directly gauges attack power, small and weak attacks are difficult to detect. 

In Figure \ref{fig:attack_filler}(b), we consdider various filler sizes ($1\%$, $3\%$, $5\%$, $10\%$, $25\%$, $40\%$) for a fixed attack size of $5\%$. We can see that altering filler size has little effect on IVD performance. Although we don't report separately, the PCA and MPE methods demonstrate almost flawless detection across diverse attack and filler sizes, corroborated by both our datasets and findings in \cite{user_cluster_1, MPE}. It's noteworthy that their detection accuracy is influenced by filler selection and rating obfuscation respectively, as elucidated in earlier sections.

Based on these findings, for our experimets, we fix an attack size of $5\%$ and filler size of $10\%$ for the rest of the analysis.

\subsection{Effect of Choice of Filler Items}

\begin{table}[htbp]
\centering
\begin{tabular}{|c|c|c|c|c|}
\hline
\multirow{2}{*}{Defense Strategy/Dataset} & \multicolumn{4}{c|}{Filler} \\
\cline{2-5}
& filler@60 & filler@40 & filler@20 & filler@10 \\
\hline
\textbf{IVD/ML-100k} &\textbf{1} &\textbf{1} &\textbf{1} &\textbf{1} \\
\hline
MPE/ML-100k &1 &1 &1 &1 \\
\hline
PCA/ML-100k &1 &0.58 &0.0 &0.0 \\
\hline
\textbf{IVD/ML-1M} &\textbf{1} &\textbf{1} &\textbf{1} &\textbf{1} \\
\hline
MPE/ML-1M &0.96 &0.97 &0.95 &0.94 \\
\hline
PCA/ML-1M &0.96 &0.31 &0.0 &0.0 \\
\hline
\textbf{IVD/Netflix} &\textbf{1} &\textbf{1} &\textbf{1} &\textbf{1} \\
\hline
MPE/Netflix &1 &1 &1 &1 \\
\hline
PCA/Netflix &1 &0.98 &0.25 & 0.0\\
\hline
\end{tabular}
\caption{Detection Rate for Different Defense Strategies and Datasets for \textbf{Average Attack} over top $x\%$ filler.  Attack size=$5\%$, Filler size=$10\%$} \label{filler_avg}
\end{table}

\begin{table}[htbp]
\centering
\begin{tabular}{|c|c|c|c|c|}
\hline
\multirow{2}{*}{Defense Strategy/Dataset} & \multicolumn{4}{c|}{Filler} \\
\cline{2-5}
& filler@60 & filler@40 & filler@20 & filler@10 \\
\hline
\textbf{IVD/ML-100k} &\textbf{0.84} &\textbf{0.85} &\textbf{0.97} &\textbf{1} \\
\hline
MPE/ML-100k &1 &1 &1 &1 \\
\hline
PCA/ML-100k &1 &0.83 &0.0 &0.0 \\
\hline
\textbf{IVD/ML-1M} &\textbf{0.87} &\textbf{1} &\textbf{1} &\textbf{1} \\
\hline
MPE/ML-1M &1 &1 &1 &1 \\
\hline
PCA/ML-1M &1 &0.63 &0.0 & 0.0\\
\hline
\textbf{IVD/Netflix} &\textbf{1} &\textbf{1} &\textbf{1} &\textbf{1} \\
\hline
MPE/Netflix &1 &1 &1 &1 \\
\hline
PCA/Netflix &1 &1 &0.43 &0.0 \\
\hline
\end{tabular}
\caption{Detection Rate for Different Defense Strategies and Datasets for \textbf{Random Attack} over top $x\%$ filler. Attack size=$5\%$, Filler size=$10\%$}\label{filler_rnd}
\end{table}

\begin{table}[htbp]
\centering
\begin{tabular}{|c|c|c|c|c|}
\hline
\multirow{2}{*}{Defense Strategy/Dataset} & \multicolumn{4}{c|}{Filler} \\
\cline{2-5}
& filler@60 & filler@40 & filler@20 & filler@10 \\
\hline
\textbf{IVD/ML-100k} &\textbf{1} &\textbf{1} &\textbf{1} &\textbf{1} \\
\hline
MPE/ML-100k &1 &1 &1 &1 \\
\hline
PCA/ML-100k &0.93 &0.58 &0.0 &0.0 \\
\hline
\textbf{IVD/ML-1M} &\textbf{1} &\textbf{1} &\textbf{1} &\textbf{1} \\
\hline
MPE/ML-1M &0.80 &0.89 &0.83 &0.88 \\
\hline
PCA/ML-1M &0.94 &0.25 &0.0 &0.0 \\
\hline
\textbf{IVD/Netflix} &\textbf{1} &\textbf{1} &\textbf{1} &\textbf{1} \\
\hline
MPE/Netflix &1 &1 &1 &1 \\
\hline
PCA/Netflix &1 &1 &0.25 &0.0 \\
\hline
\end{tabular}
\caption{Detection Rate for Different Defense Strategies and Datasets for \textbf{Target Attack} over top $x\%$ filler. Attack size=$5\%$, Filler size=$10\%$}\label{filler_tar}
\end{table}

Recall our discussion that the PCA defense strategy relies on the assumption that attackers are more similar to each other than to the genuine user base.  We choose filler items with equal probability from the top x\% of most popular items. We report detection rate for the three methods against reducing value of $x$ from the top $60\%$ to the top $10\%$ of the most popular items and compare the performance. Table \ref{filler_avg}, \ref{filler_rnd} and \ref{filler_tar} report the results for Average, Random and Target Attack respectively.

As expected, the PCA accuracy decreased significantly with increasing $x\%$ due to the increase in similarity among attack profiles and genuine profiles. As demomstrated earlier, detection becomes hopeless for $x$ values less than $20\%$ for all datasets. When filler items are selected from the popular items, their preferences become more diverse and closer to other genuine users who also rated these popular items and thus are less likely to form a well-defined cluster. This leads to a reduced accuracy of PCA as it fails to effectively distinguish attackers from genuine users.

However, both the IVD and MPE strategies still demonstrate high accuracy in comparison to PCA as discussed previously. This is because these detection methods are not reliant on the correlation between fake profiles. Specifically, MPE focuses on the deviation from the point of view of item predicted ratings rather than user profiles. As a result, it is not significantly impacted by the choice of filler items, making it more robust against variations in filler selection. Similarly IVD focuses on the deviation of the item feature vector to new ratings.  Fake user profiles have no direct impact on the item vector deviation calculation \ref{V_iterative}. Rather than working with the profiles directly, IVD employs user vector $U$ of users who provide ratings to compute the updated item vector.

\subsection{Effect of Target Shifting Obfuscation}
In this section, we will look at the effects of using target-shifting obfuscation in attack techniques. For the experiment, we fix the filler selection to the top $20\%$ of the items in the datasets. The detection rates are shown in Table \ref{obf}. 

In terms of shilling detection rate, the IVD technique surpasses both MPE and PCA. As previously noted in the previous sections, the reduced performance of PCA is primarily attributable to the selection of filler items. In contrast, MPE performance suffers because, when fraudulent users use target rating obfuscation, their MPE values may closely mirror those of real users, making differentiation difficult as demonstrated in section \ref{discussion}.  However, because IVD takes into account both the user vector of raters and the rating deviation, it is more sensitive to changes from attacks even after target-rating obfuscation.

\begin{table}[h]
\centering
\begin{tabular}{|c|c|c|c|}
\hline
\multirow{2}{*}{Defense Strategy/Dataset} & \multicolumn{3}{c|}{Attack Type} \\
\cline{2-4}
& Average & Random & Target \\
\hline
\textbf{IVD/ML-100k} &\textbf{0.86} &\textbf{0.65} & \textbf{0.90} \\
\hline
MPE/ML-100k &0.51 &0.19 &0.67 \\
\hline
PCA/ML-100k &0.0 &0.0 &0.0 \\
\hline
\textbf{IVD/ML-1M} &\textbf{0.86} &\textbf{0.60} &\textbf{0.83} \\
\hline
MPE/ML-1M &0.55 &0.29 &0.42 \\
\hline
PCA/ML-1M &0.0 &0.0 &0.0 \\
\hline
\textbf{IVD/Netflix} &\textbf{1} &\textbf{0.93} &\textbf{1} \\
\hline
MPE/Netflix &0.40 & 0.03 &0.84 \\
\hline
PCA/Netflix &0.22 &0.40 &0.25 \\
\hline
\end{tabular}
\caption{Results for Different Defense Strategies and Datasets under \textbf{Target Rating Obfuscation+Filler@20 Obfuscation}}\label{obf}
\end{table}

\subsection{Reciever Operater Characterstics}

\begin{figure}[H]
    \centering

    \begin{subfigure}[b]{0.45\textwidth}
        \centering
        \includegraphics[width=\textwidth]{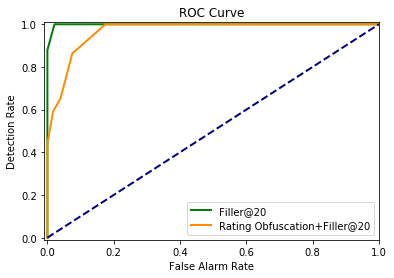}
        \caption{ML 100k: Average Attack}
    \end{subfigure}
    \hfill
    \begin{subfigure}[b]{0.45\textwidth}
        \centering
        \includegraphics[width=\textwidth]{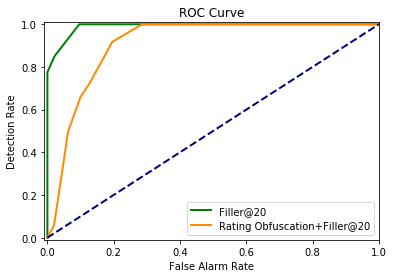}
        \caption{ML 100k: Random Attack}
    \end{subfigure}

    \vspace{1em}

    \begin{subfigure}[b]{0.45\textwidth}
        \centering
        \includegraphics[width=\textwidth]{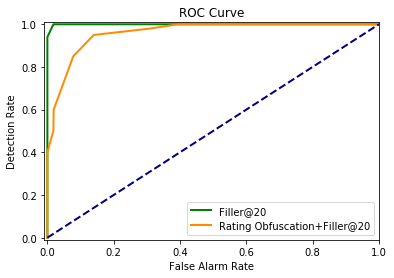}
        \caption{ML 1M: Average Attack}
    \end{subfigure}
    \hfill
    \begin{subfigure}[b]{0.45\textwidth}
        \centering
        \includegraphics[width=\textwidth]{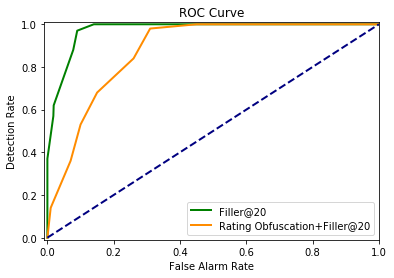}
        \caption{ML 1M: Random Attack}
    \end{subfigure}

    \caption{Reciever Operating Characterstics for ML100k and ML-1M Datasets}\label{ROC}
\end{figure}

The Receiver Operator Characteristic for the IVD method applied to two standard attack strategies: Average and Random is depicted in Figure \ref{ROC}. We simulate genuine ratings by drawing ratings from the empirical distribution of ratings for a group. When we need a rating for the target item that the user has not rated, pick a second user from the same group who has rated the item and merge the pair of user ratings.

IVD technique is applied to attack scenarios with fixed parameters: the top $20\%$ popular items as filler options, a filler size of $10\%$, and an attack size of $5\%$. For each Average and Random attack case, we examine the ROC curve for the 1) maximum target item rating model and 2) target rating obfuscated model.

For the un-obfuscated Average and Random attacks, we can see that IVD obtains a $100\%$ detection rate with a false-alarm rate of less than $10\%$ for both datasets.  

When we apply rating obfuscation to these attack strategies, we find that detection is achieved at $80\%$  for the average attack and at $~60\%$ for the random attack for the same false alarm rate of less than $10\%$. For the obfuscated attacks in both datasets, a near perfect detection rate happens with a cost of approximately $35\%$ false alarm rate. We can see that IVD performs better at detecting Obfuscated Average attacks than Random attacks. This is to be expected given that IVD directly gauges attack power. Obfuscated Random attacks are more difficult to detect since they are simpler and less effective at shifting the item vector.

\subsection{Limitations}
\begin{itemize}
    \item Cold Items:  The proposed method assumes that initial ratings for new items can be obtained in the RS.  This is a reasonable assumption since, ratings for such items may be obtained from trusted sources, such as professional critics, trusted representatives \cite{repr_based_cold_start_1, repr_based_cold_start_2}, or filter bots \cite{filter_bots}.  We could also randomly elicit ratings from users in the system to reduce the risk of sampling fake ratings from false users already existing in RS. For instance, Goodreads  \footnote{\url{https://help.goodreads.com/s/announcements/a031H00000RKE8VQAX/giveaways-for-authors-frequently-asked-questions?ref=kdpnl}} and 'Voracious readers only' \footnote{\url{https://voraciousreadersonly.com/}} give away free copies of new books to readers in exchange for voluntary ratings. At the same time, movie RS offer rewards for watching new movies and TV series and provide initial reviews \footnote{\url{https://www.amazon.in/b/ref=as_li_ss_tl?node=15697217031&ref_=dvm_crs_merch_in_ai_slashpv_P_watch&wint2_header&pf_rd_m=A1K21FY43GMZF8&pf_rd_s=merchandised-search-2&pf_rd_r=4N0T0PVXWA9S566NFGQV&pf_rd_t=101&pf_rd_p=51f6c505-5ad6-4b9c-b9b6-91450c336884&pf_rd_i=10882806031&linkCode=sl2&tag=thinkerviewsc-21&linkId=eb862e43198dfa2f8e5a415bb4db6f26&language=en_IN}}. We further assume that the randomly selected users are real and that the likelihood of all initial users and ratings being fake is low. We note that the RS does not have to wait for $20-25$ ratings per group before recommending the new item to existing system users. We simply propose that the RS avoid using unprompted ratings for the new item in the training process until it receives ratings from reliable sources upto the threshold.

    \item A reasonable choice of reference and threshold are crucial in determining the performance of the proposed approach. We demonstrated the detection in both a dense dataset and two complete datasets. IVD results in reasonable performance in all the cases.
    In practice, utilizing a single reference cluster as a broad assumption may yield insights and preliminary findings. However, in order to create a viable fake ratings detection mechanism, additional techniques such as dynamic references, weighted references, or models that learn references adaptively over time can be explored as part of future work. These methods may better capture the intricacies of user preferences, deal with potential changes, and accommodate a wide range of scenarios, resulting in increased detection accuracy.
    
\end{itemize}

\subsection{Conclusions}
We present a new approach for detecting shilling attacks that makes use of item preference vectors. The Item Vector Deviation (IVD) technique provides an unsupervised and attack model-free strategy that can be deployed directly to any target item, provided the recommender system has sufficient rating information about the item in question to compute its initial preference vector. Remarkably, as low as $20-25$ ratings per user cluster were sufficient to produce favorable outcomes across all examined datasets.

Our experiments confirm the sustained effectiveness of our proposed strategy, achieving high detection accuracy with minimal false alarms. This performance is stable across a wide range of attack scenarios, and it even performs reasonably well against obfuscated attacks, outperforming our baselines. 
 
\begin{acks}
This work was supported by Science Foundation Ireland (SFI) under grant 16/IA/4610
\end{acks}

\bibliographystyle{ACM-Reference-Format}
\bibliography{ivd}

\appendix
\section{APPENDIX}
The proof and discussion below are taken from \cite{TORS}.

Given $M$ is a square $n \times n$ matrix whose inverse matrix we know, for the simple case of a rank-1 perturbation to $M$, Sherman–Morrison formula provides a method to find the updated rank-1 change to the inverse.

\begin{Definition}\label{sherman_morrison}
 Given $M$ is a square $n \times n$ matrix whose inverse matrix we know, $u$ and $v$ are $n \times 1$ column vectors defining the perturbation to matrix $M$ such that $u_iv_j^T$ is added to $M_{i,j}$, then we can find the inverse of the modified $M$ by Sherman-Morrison formula
 
 $$(M+uv^T)^{-1}=M^{-1}-\frac{M^{-1}uv^TM^{-1}}{1+v^TM^{-1}u}$$
\end{Definition}
\noindent For a generalised rank-$k$ perturbation to $M$, the updated inverse is given by the following formula.
\begin{Definition}\label{woodbury_identity}
 Given $M$ is a square $n \times n$ matrix whose inverse matrix we know, $U$ and $V$ are $n \times k$ matrices defining rank $k$ perturbation to matrix $M$, then we can find the inverse of the modified $M$ 
 
 $$(M+UV^T)^{-1}=M^{-1}-M^{-1}U(I_k+V^TM^{-1}U)^{-1}V^TM^{-1}$$
\end{Definition}

\subsection{Proof of Theorem \ref{V_iterative}}
\begin{Proof}
From equation \ref{costfunc_recap}, given last updated $U$, $V_{j^*}$ of item $j^*$ is given by: 
\begin{align} \label{Uconst}
V_{j^*}&=\left(\sum_{i \in \mathcal{U}(j^*)}{U_iU_i^T+\lambda I}\right)^{-1}\sum_{i \in \mathcal{U}(j^*)}U_iR_{i,j^*} 
\end{align}
where $\mathcal{U}(j^*)=\{i:(i,j^*) \in \mathcal{O}\}$ is the set of users rating item $j^*$. For simplicity, let $\Lambda$ gather together the feature vectors $U_i$ of all users $i$ who rated target item $j^*$ such that $\Lambda\Lambda^T=\sum_{i \in \mathcal{U}(j^*)}U_iU_i^T$ and let $r$ be a column vector of ratings received by the target item from these true users. 
We can write equation \ref{Uconst} as,
$$V_{j^*}=(\Lambda\Lambda^T+\lambda I)^{-1}\Lambda r=A^{-1}\Lambda r$$  where $A=\Lambda\Lambda^T+\lambda I$. 

After the block of new users enters with feature vectors represented by the columns of $X$, we write $\hat{U}=[\Lambda,X]$ and $\hat{U}\hat{U}^T=\Lambda\Lambda^T+XX^T$. Also $\hat{r}=\begin{bmatrix}
r  \\
y  
\end{bmatrix}
$
and
$\hat{A}^{-1}=(\hat{U}\hat{U}^T+\lambda I)^{-1}=(A+XX^T)^{-1}$. We now calculate the updated $\hat{V_{j^*}}$ after attack. We know that, 
\begin{align*}
 \hat{V_{j^*}}&= \hat{A}^{-1}\hat{U}\hat{r} = \hat{A}^{-1}\left(\Lambda r+Xy\right)= \left(A+XX^T\right)^{-1}\left(\Lambda r+Xy\right) \\
        &\text{Applying Sherman Morrison/Woodbury formula to the inverse (see Definition \ref{woodbury_identity})}\\
        &= \left(A^{-1}-A^{-1}X\left(I+X^TA^{-1}X\right)^{-1}X^TA^{-1}\right) \left(\Lambda r+Xy\right)\\
        &= A^{-1}\Lambda r-A^{-1}X\left(I+X^TA^{-1}X\right)^{-1}X^TA^{-1}\Lambda r+A^{-1}Xy-A^{-1}X\left(I+X^TA^{-1}X\right)^{-1}X^TA^{-1}Xy \\
        &\text{We have, } V_{j^*}=A^{-1}\Lambda r \text{. substituting for $A^{-1}\Lambda r$ and rearranging, we get, }\\
        &= V_{j^*}-A^{-1}X\left(I+X^TA^{-1}X\right)^{-1}X^TV_{j^*}+A^{-1}X\left(I- \left(I+X^TA^{-1}X\right)^{-1}X^TA^{-1}X\right)y \\
        &\text{Substituting } I=(I+X^TA^{-1}X)^{-1}(I+X^TA^{-1}X) \text{ inside third term, we get,}\\
        \begin{split}
        &= V_{j^*}-A^{-1}X\left(I+X^TA^{-1}X\right)^{-1}X^TV_{j^*}+ \\ &A^{-1}X\left(\left(I+X^TA^{-1}X\right)^{-1}\left(I+X^TA^{-1}X\right)-\left(I+X^TA^{-1}X\right)^{-1}X^TA^{-1}X\right)y 
        \end{split}\\
        &= V_{j^*}-A^{-1}X\left(I+X^TA^{-1}X\right)^{-1}X^TV_{j^*}
        +A^{-1}X\left(\left(I+X^TA^{-1}X\right)^{-1}\left(I+X^TA^{-1}X-X^TA^{-1}X\right)\right)y\\
       &=V_{j^*}-A^{-1}X\left(I+X^TA^{-1}X\right)^{-1}X^TV_{j^*}+A^{-1}X\left(I+X^TA^{-1}X\right)^{-1}y\\
       &=V_{j^*}+A^{-1}X\left(I+X^TA^{-1}X\right)^{-1}\left(y-X^TV_{j^*}\right)
\end{align*}
\end{Proof}

Finally, we have,
\begin{align} \label{Uconst_iter}
\hat{V_{j^*}}-V_{j^*}&=A^{-1}X\left(I+X^TA^{-1}X\right)^{-1}\left(y-X^TV_{j^*}\right)
\end{align}

\end{document}